\documentclass[3p]{elsarticle}

\usepackage{amsmath,amssymb}
\usepackage{graphicx}

\begin{document}

\begin{frontmatter}

\title{Spreading lengths of Hermite polynomials}

\author[ma,cp]{P. S\'anchez-Moreno}
\ead{pablos@ugr.es}
\author[fm,cp]{J.S. Dehesa}
\ead{dehesa@ugr.es}
\author[fm,cp]{D. Manzano}
\ead{manzano@ugr.es}
\author[ma,cp]{R.J. Y\'a\~nez}
\ead{ryanez@ugr.es}

\address[ma]{Departamento de Matem\'atica Aplicada, Universidad de Granada, Granada, Spain}
\address[fm]{Departamento de F\'{\i}sica At\'omica, Molecular y Nuclear, Universidad de Granada, Granada, Spain}
\address[cp]{Instituto ``Carlos I'' de F\'{\i}sica Te\'orica y Computacional, Universidad de Granada, Granada, Spain}

\begin{abstract}
The Renyi, Shannon and Fisher spreading lengths of the classical or hypergeometric orthogonal polynomials, which are quantifiers of their distribution all over the orthogonality interval, are defined and investigated. These information-theoretic measures of the associated Rakhmanov probability density, which are direct measures of the polynomial spreading in the sense of having the same units as the variable, share interesting properties: invariance under translations and reflections, linear scaling and vanishing in the limit that the variable tends towards a given definite value. The expressions of the Renyi and Fisher lengths for the Hermite polynomials are computed in terms of the polynomial degree. The combinatorial multivariable Bell polynomials, which are shown to characterize the finite power of an arbitrary polynomial, play a relevant role for the computation of these information-theoretic lengths. Indeed these polynomials allow us to design an error-free computing approach for the entropic moments (weighted $L^q$-norms) of Hermite polynomials and subsequently for the Renyi and Tsallis entropies, as well as for the Renyi spreading lengths. Sharp bounds for the Shannon length of these polynomials are also given by means of an information-theoretic-based optimization procedure. Moreover, it is computationally proved the existence of a linear correlation between the Shannon length (as well as the second-order Renyi length) and the standard deviation. Finally, the application to the most popular quantum-mechanical prototype system, the harmonic oscillator, is discussed and some relevant asymptotical open issues related to the entropic moments mentioned previously are posed.
\end{abstract}


\end{frontmatter}

\section{Introduction}
Let $\{ p_n(x) \}$ denote a sequence of real orthonormal polynomials with respect to the weight function $\omega (x)$ on the interval $\Delta \equiv (a,b)\subseteq \mathbb{R}$ (see e.g. \cite{temme_96,nikiforov_88}), i.e.
\begin{equation}
\int_{\Delta} p_n(x) p_m(x) \omega(x) dx=\delta_{n,m}, \quad m,n \in \mathbb{N}
\label{eq:orthonormality}
\end{equation}

The distribution of these polynomials along the orthogonality interval can be complementarily measured by means of the spreading properties of the normalized-to-unity density function
\begin{equation}
\rho_n(x)\equiv\rho[p_n]=p_n^2 (x) \omega(x),
\label{eq:rakhmanov_density}
\end{equation}
which is called Rakhmanov's density of the polynomial $p_n(x)$, to honor the pioneering work \cite{rakhmanov:mus77} of this mathematician who has shown that this density governs the asymptotic $(n\to \infty)$ behaviour of the ratio $p_{n+1}/p_n$. Physically, this probability density characterizes the stationary states of a large class of quantum-mechanical potentials \cite{nikiforov_88}.
Beyond the variance, the spreading of the orthogonal polynomials is best analyzed by the information-theoretic properties of their associated Rakhmanov probability densities; namely, the Fisher information \cite{fisher:pcps25},  the Renyi entropy \cite{renyi_70} and the  Shannon entropy \cite{shannon_49}.

The information-theoretic knowledge of the orthogonal polynomials up to 2001 is reviewed in Ref \cite{dehesa:jcam01}, where the quantum-mechanical motivation and some physical applications are also given, and up today in \cite{aptekarev:jcam09}, where emphasis is made on asymptotics. Therein, it is pointed out that the study of the information-theoretic measures of orthogonal polynomials was initiated in the nineties with the asymptotic computation of the Renyi and Shannon entropies of the classical orthogonal polynomials \cite{aptekarev:rassm95,dehesa:pra94,dehesa:maa97}. Up until now, however, the explicit expressions of these two spreading measures are not known save for the Shannon measure for some particular subclasses of the Jacobi polynomials; namely the Chebyshev polynomials of first and second type \cite{dehesa:pra94,dehesa:maa97} and the Gegenbauer polynomials $C_n^\lambda(x)$ with integer parameter $\lambda$ \cite{buyarov:jpa00,devicente:jpa07}. On the other hand, the Fisher information for all classical orthogonal polynomials on a real interval (Hermite, Laguerre, Jacobi) has been calculated in a closed form in 2005 \cite{sanchezruiz:jcam05}. This has allowed us to find, more recently, the Cramer-Rao information plane (i.e. the plane defined by both the Fisher information and the variance) for these systems \cite{dehesa:jcam06}.

This paper has two main purposes. First, to introduce a more appropriate set of spreading measures (to be called spreading lengths of Renyi, Shannon and Fisher types) in the field of orthogonal polynomials following the ideas of M.J.W. Hall \cite{hall:pra99}. The spreading lengths are direct measures of position spread of the variable $x$, as the root-mean-square or standard deviation $\Delta x$. So they have the following properties: same units as $x$, invariance under translations and reflections, linear scaling with $x$ and vanishing in the limit as $\rho_n(x)$ approaches a delta function. 
The second purpose is to compute all these lengths for the Hermite polynomials. In this sense we first compute their moments around the origin and, as well, the standard deviation of these objects. Then we compute the entropic or frequency moments \cite{shenton:b51,sichel:b49,sichel:jrss47} of these polynomials by means of the Bell polynomials, so useful in Combinatorics \cite{comtet_74,godsil:c82}. This achievement allows us to obtain the Renyi lengths of Hermite polynomials by means of an efficient algorithm based on the known properties of Bell's polynomials. With respect to the Shannon length, since its explicit expression cannot be determined, we obtain its asymptotics and we derive a family of sharp upper bounds by use of an information-theoretic-based optimization procedure. Moreover, the Fisher information is explicitly calculated. Finally, these information-theoretic quantities are applied to the computation of the spreading lengths of the main quantum-mechanical prototype of physical systems: the harmonic oscillator.

The structure of the paper is the following. In Section \ref{sec2} we describe the spreading lengths for the orthogonal polynomials $\{p_n(x)\}$ and discuss their properties. 
In Section \ref{sec3} the computational methodology to obtain the entropic moments and the associated Renyi lengths of the Hermite polynomials is developed by use of the Bell polynomials. In addition, the moments with respect to the origin and the Fisher length are explicitly given. Furthermore, the asymptotics of the Hermite polynomials is obtained and a family of sharp bounds are variationally determined. In Section \ref{sec4} the spreading lengths of Hermite polynomials is numerically discussed and a linear correlation with the standard deviation is computationally discovered. In Section \ref{sec5} the previous results are applied to the quantum-mechanical harmonic oscillator. Finally, some open problems about the asymptotics of entropic moments are posed, our conclusions are given and an Appendix with the computational approach of the power of an arbitrary polynomial by use of the Bell polynomials is included.

\section{Spreading lengths of classical orthogonal polynomials: Basics}
\label{sec2}

Here we shall refer to the hypergeometric-type polynomials, i.e. the polynomials $\{p_n(x)\}$ defined by the orthogonality relation given by Eq. (\ref{eq:orthonormality}). It is well-known that they can be reduced to one of the three classical families of Hermite, Jacobi and Laguerre by appropriate linear changes of the variable \cite{nikiforov_88}. The spreading of these polynomials along their orthogonality interval $\Delta\equiv (a,b)$ can be measured by means of (i) the moments around a particular point of the orthogonality interval; usually it is chosen the origin (moments around the origin: $\mu_k'=\left< x^k\right> _n$) or the centroid $\left< x\right>$ (central moments or moments around the centroid: $\mu_k=\langle(x-\langle x\rangle_n)^k\rangle_n $) \cite{kendall_69} and (ii) the frequency moments $W_k[\rho_n]:=\langle \left[ \rho_n(x)\right] ^k\rangle$ \cite{yule:b38,sichel:jrss47,kendall_69} of the associated Rakhmanov density $\rho_n(x)$ given by Eq. (\ref{eq:rakhmanov_density}). The symbol $\left< f(x)\right>\equiv \left< f(x)\right> _n$ denotes the expectation value of $f(x)$ with respect to the density $\rho_n(x)$ as
\[
\left< f(x)\right>_n := \int_{\Delta} f(x)\rho_n(x)dx=\int_\Delta f(x)\omega(x) p_n^2(x) dx.
\]

These two classes of ordinary and frequency moments are complementary spreading measures of global type. While the ordinary moments measures the distribution of the probability with respect to a particular point of the support interval $\Delta$, the frequency moments measure the extent to which the probability is in fact distributed. So it happens that the latter moments are, at times, much better probability estimators than the ordinary ones \cite{sichel:jrss47,sichel:b49}. Moreover, they are fairly efficient in the range where the ordinary moments are fairly inefficient  \cite{shenton:b51}; see also the brief, recent summary about these quantities done in Ref. \cite{romera_02}. Therein we learn that the frequency moments are also called "entropic moments" because they are closely connected to the Renyi entropies \cite{renyi_70}
\begin{equation}
R_q[\rho_n]:=\frac{1}{1-q}\ln \langle [\rho_n(x)]^{q-1}\rangle;\quad q>0;\quad q\neq 1,
\label{eq:def_renyi}
\end{equation}
and the Tsallis entropies \cite{tsallis:jsp88}
\begin{equation}
T_q[\rho_n]:=\frac{1}{q-1}\left[ 1-\langle [\rho_n(x)]^{q-1}\rangle\right];\quad q>0; \quad q\neq 1, 
\label{eq:def_tsallis}
\end{equation}
as well as to other information-theoretic quantities, such as the Brukner-Zeilinger entropy \cite{brukner:pra01} and the linear entropy \cite{zurek:prl93}.

Among the first class of moments we highlight the familiar root-mean-square or standard deviation $(\Delta x)_n$ given by
\[
(\Delta x)_n = \left( \left< x^2\right>_n -\left< x\right>_n^2\right)^{\frac{1}{2}},
\]
because it is a direct measure of spreading \cite{hall:pra99} in the sense of having the same units as the variable, and has various interesting properties: invariance under translations and reflections, linear scaling (i.e., $\Delta y=\lambda\Delta x$ for $y=\lambda x$) and vanishing as the density approaches a Dirac delta density (i.e. in the limit that $x$ tends towards a given definite value).

From the second class of moments we fix our attention on the Renyi lengths \cite{hall:pra99} defined by 
\begin{equation}
L_q^R[\rho_n]\equiv \exp\left(R_q[\rho_n]\right) = \langle \left[\rho_n(x)\right]^{q-1}\rangle^{-\frac{1}{q-1}}=\left\{\int_{\Delta}dx[\rho_n(x)]^{q} \right\}^{-\frac{1}{q-1}},\quad q>0, \quad q\neq 1,
\label{eq:def_renyi_length}
\end{equation}
which are not only direct spreading measures but also have the three above-mentioned properties of the standard deviation. Among them we highlight the second-order Renyi length (also called Onicescu information \cite{onicescu:craspa66}, Heller length \cite{heller:pra87}, disequilibrium \cite{anteneodo:pla96,lopez:pla95} and inverse participation ratio in other contexts \cite{zyckwski:jpa90,mirbach:ap98})

\[
L[\rho_n]:= L_2^R[\rho_n]=\langle \rho_n(x)\rangle^{-1}=\left\{ \int_{\Delta} dx [\rho_n(x)]^2\right\}^{-1},
\]
and, above all, the Shannon length \cite{shannon_49,hall:pra99}
\[
N[\rho_n]:= \lim_{q\to 1} L_q^R[\rho_n]=\exp\left( S[\rho_n]\right) \equiv \exp \left[-\int_{\Delta} dx \rho_n(x) \ln\rho_n(x)\right],
\]
where $S[\rho_n]$ is the Shannon information entropy \cite{shannon_49}.

There exists a third class of spreading measures for the classical orthogonal polynomials, which are qualitatively different in the sense that, in contrast to the previous ones, they have a local character. Indeed they are functionals of the derivative of the associated Rakhmanov density $\rho_n(x)$, so that they are very sensitive to local rearrangements of the variable. The most distinctive measure of this class is the so-called Fisher length  \cite{fisher:pcps25,hall:pra01,hall:jpa02} defined by
\begin{equation}
(\delta x)_n:=\frac{1}{\sqrt{F[\rho_n]}}\equiv \left< \left[ \frac{d}{dx}\ln\rho_n(x)\right]^2\right>^{-\frac{1}{2}}=\left\{\int_{\Delta}dx\frac{[\rho_n'(x)]^2}{\rho_n(x)}\right\}^{-\frac{1}{2}},
\label{eq:def_deltax}
\end{equation}
where $F[\rho_n]$ denotes the Fisher information of the classical orthogonal polynomials \cite{sanchezruiz:jcam05}. 
This quantity measures the pointwise concentration of the probability along the orthogonality interval, and quantifies the gradient content of the Rakhmanov density providing (i) a quantitative estimation of the oscillatory character of the density and of the polynomials and (ii) the bias to particular points of the interval, so that it measures the degree of local disorder.

It is worthy to remark that the Fisher length, as the Heisenberg and Renyi lengths, is a direct spreading measure and has the three properties of translation and reflection invariance, linear scaling and vanishing when the density tends to a delta density. In addition, the Fisher length is finite for all distributions \cite{hall:pra01}. Moreover, the direct spreading measures just mentioned have an uncertainty/certainty property: see Refs. \cite{hall:pra99} for the Heisenberg, Shannon and Renyi cases, and \cite{dehesa:mp06,dehesa:jpa07} for the Fisher case. Finally, they fulfill the inequalities \cite{hall:pra99} 
\begin{equation}
(\delta x)_n\leq (\Delta x)_n, \quad \text{and} \quad N[\rho_n]\leq (2\pi e)^{\frac{1}{2}}(\Delta x)_n,
\label{eq:delta_inequalities}
\end{equation}
where the equality is reached if and only if $\rho_n(x)$ is a Gaussian density.

\section{Spreading lengths of Hermite polynomials}
\label{sec3}

In this section we calculate the moments-with-respect-to-the-origin $\langle x^k\rangle$, $k\in\mathbb{Z}$, the entropic moments $W_q[\rho_n]\equiv \langle[\rho_n(x)]^q\rangle$, of integer order $q$, and the Renyi, Shannon and Fisher spreading lengths of the Rakhmanov density of the orthonormal Hermite polynomials $\tilde{H}_n(x)$ given by
\begin{equation}
\tilde{H}_n(x)=\sum_{l=0}^n c_l x^l,
\label{eq:def_H_alternative}
\end{equation}
with
\begin{equation}
c_l=\frac{(-1)^{\frac{3n-l}{2}}n!}{\left(2^n n!\sqrt{\pi}\right)^\frac12}\frac{2^l}{\left(\frac{n-l}{2}\right)!l!}\frac{(-1)^l+(-1)^n}{2},
\label{eq:def_c_l}
\end{equation}
where the last factor vanishes when $l$ and $n$ have opposite parities. These polynomials are known to fulfil the orthonormality relation (\ref{eq:orthonormality}) with the weight function $\omega_H(x)=e^{-x^2}$ on the whole real line \cite{temme_96,nikiforov_88}. Thus, according to Eq. (\ref{eq:rakhmanov_density}), the expression
\begin{equation}
\rho_n(x)\equiv\rho[\tilde{H}_n](x)=e^{-x^2} \tilde{H}_n^2(x)
\label{eq:rho_n_hermite}
\end{equation}
gives the Rakhmanov density of the orthonormal Hermite polynomial $\tilde{H}_n(x)$.

\subsection{Moments with respect to the origin and the standard deviation}

The moments of the Hermite polynomials $\tilde{H}_n(x)$ are defined by the moments of its associated Rakhmanov density (\ref{eq:rakhmanov_density}); that is
\begin{equation}
\langle x^k\rangle_n:=\int_{-\infty}^{\infty} x^k \tilde{H}_n^2(x)e^{-x^2}dx, \; k=0,1,2,\ldots
\label{eq:def_xk}
\end{equation}
The use of (\ref{eq:def_H_alternative}), (\ref{eq:orthonormality}) and (\ref{eq:def_xk}) gives
\begin{eqnarray}
\langle x^k\rangle_n=\left\{
\begin{array}{ll}
\frac{k!}{2^k \Gamma\left(\frac{k}{2}+1\right)}\,_2F_1\left(\left.
\begin{array}{c}
-n,-\frac{k}{2}\\
1
\end{array}
\right| 2
\right),
& \text{even } k\\
0, & \text{odd }k
\end{array}
\right.,
\label{eq:xk}
\end{eqnarray}
which can be also obtained from some tabulated special integrals \cite{prudnikov_86}.

In the particular cases $k=0,1$ and $2$ we obtain the first few moments around the origin:
\[
\langle x^0\rangle_n=1,\;\langle x\rangle_n =0,\; \langle x^2\rangle_n=n+\frac12,
\]
so that the second central moment of the density
\begin{equation}
(\Delta x)_n=\sqrt{\langle x^2\rangle_n-\langle x\rangle_n^2}=\sqrt{n+\frac12}
\label{eq:Delta_x}
\end{equation}
describes the standard deviation.

\subsection{Entropic moments and Renyi lengths of integer order $q$}

The $q$th-order frequency or entropic moment of the Hermite polynomials $\tilde{H}_n(x)$ is defined by the corresponding quantity of its associated Rakhmanov density given by Eq. (\ref{eq:rakhmanov_density}); that is,
\begin{equation}
W_q[\rho_n]=\left\langle \left\{\rho_n\right\}^{q-1}\right\rangle=\int_{-\infty}^\infty \left\{\rho_n(x)\right\}^q dx =\int_{-\infty}^\infty e^{-qx^2}\{\tilde{H}_n(x)\}^{2q}dx;\; q\ge 1
\label{eq:def_wq_hermite}
\end{equation}

The evaluation of this quantity in a closed form is not at all a trivial task despite numerous efforts published in the literature \cite{azor:sjma82,godsil:c82,desainte:lnm85} save for some special cases. Here we will do it by means of the Bell polynomials which play a relevant role in Combinatorics \cite{lavoie:sjam75,riordan_80}. We start from the explicit expression (\ref{eq:def_H_alternative}) for the Hermite polynomial $\tilde{H}_n(x)$; then, taking into account the Appendix below we have that its $p$th-power can be written down as
\begin{equation}
[\tilde{H}_n(x)]^p=\left[\sum_{k=0}^n c_kx^k\right]^p =\sum_{k=0}^{np} \frac{p!}{(k+p)!}B_{k+p,p}(c_0,2!c_1,\ldots,(k+1)!c_k) x^k,
\label{eq:Hnp_bell_expanssion}
\end{equation}
with $c_i=0$ for $i>n$,
and the remaining coefficients are given by Eq. (\ref{eq:def_c_l}). Moreover, the Bell polynomials are given by
\begin{equation}
B_{m,l}(c_1,c_2,\ldots,c_{m-l+1})=\sum_{\hat{\pi}(m,l)} \frac{m!}{j_1! j_2!\cdots j_{m-l+1}!}\left(\frac{c_1}{1!}\right)^{j_1} \left(\frac{c_2}{2!}\right)^{j_2}\cdots \left(\frac{c_{m-l+1}}{(m-l+1)!}\right)^{j_{m-l+1}}
\label{eq:bell_def}
\end{equation}
where the sum runs over all partitions $\hat{\pi}(m,l)$ such that
\begin{equation}
j_1+j_2+\cdots+j_{m-l+1}=l, \quad \text{and}\quad j_1+2j_2+\cdots+(m-l+1)j_{m-l+1}=m.
\label{eq:partitions}
\end{equation}

The substitution of the expression (\ref{eq:Hnp_bell_expanssion}) with $p=2q$ into Eq. (\ref{eq:def_wq_hermite}) yields the value
\begin{eqnarray}
W_q[\rho_n]&=&\sum_{k=0}^{2nq} \frac{(2q)!}{(k+2q)!}B_{k+2q,2q}(c_0,2!c_1,\ldots,(k+1)!c_k)
\int_{-\infty}^\infty e^{-qx^2}x^kdx\nonumber\\
&=&\sum_{j=0}^{nq}\frac{\Gamma\left(j+\frac12\right)}{q^{j+\frac12}}\frac{(2q)!}{(2j+2q)!}
B_{2j+2q,2q}(c_0,2!c_1,\ldots,(2j+1)!c_{2j}),
\label{eq:wqrhon_bell}
\end{eqnarray}
where the parameters $c_i$ are given by Eq. (\ref{eq:def_c_l}), keeping in mind that $c_i=0$ for every $i>n$, so that the only non-vanishing terms correspond to those with $j_{i+1}=0$ so that $c_i^{j_{i+1}}=1$ for every $i>n$. In the particular cases $q=1$ and 2, we obtain the value
\[
W_1[\rho_n]=\int_{-\infty}^\infty \rho_n(x)dx=1
\]
for the normalization of $\rho_n(x)$, as one would expect, and
\begin{eqnarray}
W_2[\rho_n]&=&\int_{-\infty}^\infty \left\{\rho_n(x)\right\}^2dx=\int_{-\infty}^\infty e^{-2x^2} (\tilde{H}_n(x))^4dx\nonumber\\
&=&\sum_{j=0}^{2n}\frac{\Gamma\left(j+\frac12\right)}{2^{j+\frac12}}\frac{4!}{(2j+4)!}
B_{2j+4,4}(c_0,2!c_1,\ldots,(2j+1)!c_{2j})
\label{eq:W_2_n}
\end{eqnarray}
for the second-order entropic moment.

As well, the entropic moments of arbitrary order $q$ for the Hermite polynomials of lowest degree (e.g., $n=0,1$ and $2$) have the values
\[
W_q[\rho_0]=\sqrt{\frac{\pi^{1-q}}{q}},\qquad
W_q[\rho_1]= \frac{2^q}{\pi^\frac{q}{2}q^{q+\frac12}} \Gamma\left(q+\frac{1}{2}\right),
\qquad
W_q[\rho_2]=\frac{\pi^\frac{1-q}{2}2^q(2q)!}{q^{2q+\frac12}}\mathcal{L}_{2q}^{(-2q-\frac12)}\left(-\frac{q}{2}\right),
\]
where $\mathcal{L}_n^{(\alpha)}(x)$ denotes a Laguerre polynomial.

Then, we can now calculate all the information-theoretic measures of the Hermite polynomials which are based on their entropic moments $W_q[\rho_n]$ just calculated, such as the Renyi and Tsallis entropies given by Eqs. (\ref{eq:def_renyi}) and (\ref{eq:def_tsallis}), respectively, and the Brukner-Zeilinger and linear entropy which are closely related to the first-order entropic moment $W_2[\tilde{H}_n]$ given by Eq. (\ref{eq:W_2_n}). Particularly relevant are the values
\begin{equation}
L_q^R[\rho_n]=\left\{W_q[\rho_n]\right\}^{-\frac{1}{q-1}}=\left(\sum_{j=0}^{nq}\frac{\Gamma\left(j+\frac12\right)}{q^{j+\frac12}}\frac{(2q)!}{(2j+2q)!}
B_{2j+2q,2q}(c_0,2!c_1,\ldots,(2j+1)!c_{2j})\right)^{-\frac{1}{q-1}};\;q=2,3,\ldots
\label{eq:lqrhon}
\end{equation}
for the Renyi lengths (\ref{eq:def_renyi_length}) of Hermite polynomials, and
\[
L[\rho_n]=\left\{W_2[\rho_n]\right\}^{-1}=\left(\sum_{j=0}^{2n}\frac{\Gamma\left(j+\frac12\right)}{2^{j+\frac12}}\frac{4!}{(2j+4)!}
B_{2j+4,4}(c_0,2!c_1,\ldots,(2j+1)!c_{2j})\right)^{-1},
\]
for the Heller length of Hermite polynomials.

Let us just explicitly write down the values
\[
L_q^R[\rho_0]=\pi^\frac12 q^\frac{1}{2(q-1)}
\]
\[
L_q^R[\rho_1]=\left(\frac{\pi^\frac12 q}{2}\right)^\frac{q}{q-1} q^\frac{1}{2(q-1)}
\left(\Gamma\left(q+\frac12\right)\right)^{-\frac{1}{q-1}}
\]
\[
L_q^R[\rho_2]=\pi^\frac12 2^\frac{q}{1-q} q^\frac{4q+1}{2q-2}\left((2q)!\mathcal{L}_{2q}^{(-2q-\frac12)}\left(-\frac{q}{2}\right)\right)^{-\frac{1}{q-1}}
\]
for the Renyi lengths of the first few Hermite polynomials with lowest degrees, and the values
\[
L[\rho_0]=\sqrt{2\pi},\qquad
L[\rho_1]=\frac43\sqrt{2\pi},\qquad
L[\rho_2]=\frac{64}{41}\sqrt{2\pi}
\]
for the Onicescu-Heller length of these polynomials.

\subsection{Shannon length: Asymptotics and sharp bounds}

Here we determine the asymptotics of the Shannon length $N[\rho_n]$ of the Hermite polynomials $\tilde{H}_n(x)$ and its relation with its standard deviation. Moreover we use an information-theoretic-based optimization procedure to find sharp upper bounds to $N[\rho_n]$ and we discuss their behavior in a numerical way.

The Shannon length or exponential entropy of the Hermite polynomials $\tilde{H}_n(x)$ is defined by
\begin{equation}
N[\rho_n]=\exp\left\{S[\rho_n]\right\},
\label{eq:nrhosrho}
\end{equation}
where
\begin{equation}
S[\rho_n]=-\int_{-\infty}^\infty e^{-x^2} \tilde{H}_n^2(x)\ln\left[e^{-x^2}\tilde{H}_n^2(x)\right]dx
\label{eq:shonnonrhon}
\end{equation}
is the Shannon entropy of the Hermite polynomial of degree $n$. Simple algebraic operations yield to
\begin{equation}
S[\rho_n]=n+\frac12-\int_{-\infty}^\infty e^{-x^2} \tilde{H}_n^2(x)\ln\tilde{H}_n^2(x)dx
\label{eq:shannon_H}
\end{equation}

The logarithmic integral involved in this expression has not yet been determined in spite of serious attempts for various authors \cite{aptekarev:dm96,aptekarev:rassm95,dehesa:pra94,dehesa:maa97}. Nevertheless these authors have found its value for large $n$ by use of the strong asymptotics of Hermite polynomials; it is given as
\begin{equation}
\int_{-\infty}^\infty e^{-x^2} \tilde{H}_n^2(x)\ln\tilde{H}_n^2(x)dx =n+\frac32-\ln\pi-\ln\sqrt{2n}+o(1).
\label{eq:shannon_HlnH}
\end{equation}
Then, from (\ref{eq:shannon_H}) and (\ref{eq:shannon_HlnH}) one has the asymptotical value
\[
S[\rho_n]=\ln\sqrt{2n}+\ln\pi-1+o(1)=\ln\frac{\pi\sqrt{2n}}{e}+o(1)
\]
for the Shannon entropy, and
\begin{equation}
N[\rho_n]\simeq \frac{\pi\sqrt{2n}}{e};\;n\gg 1
\label{eq:N_H}
\end{equation}
for the Shannon length of the Hermite polynomial $\tilde{H}_n(x)$. From Eqs. (\ref{eq:Delta_x}) and (\ref{eq:N_H}) one finds the following relation
\begin{equation}
N[\rho_n]\simeq \frac{\pi\sqrt{2}}{e} (\Delta x)_n;\;n\gg 1
\label{eq:nrho_delta}
\end{equation}
between the asymptotical values of the Shannon length and the standard deviation of the Hermite polynomial $\tilde{H}_n$ \cite{devicente_04}. Remark that this relation fulfils the general inequality (\ref{eq:delta_inequalities}) which mutually relate the Shannon length and the standard deviation.

Since the evaluation of the Shannon $S(\tilde{H}_n)$ given by Eq. (\ref{eq:shannon_H}) is not yet possible for a generic degree, it seems natural to try to find sharp bounds to this quantity. Here we do that by use of the non-negativity of the Kullback-Leibler entropy of the two arbitrary probability densities $\rho(x)$ and $f(x)$ defined by
\[
I_{\rm KL}[\rho,f]=\int_{-\infty}^\infty \rho(x)\ln\frac{\rho(x)}{f(x)}dx.
\]
Indeed, for $\rho(x)=\rho_n(x)$, the Rakhmanov density (\ref{eq:rakhmanov_density}) associated to the Hermite polynomials, the non-negativity of the corresponding Kullback-Leibler functional yields the following upper bound to $S(\tilde{H}_n)$:
\begin{equation}
S[\rho_n]\le -\int_{-\infty}^\infty \rho_n(x)\ln f(x)dx,
\label{eq:S_le_rholnf}
\end{equation}
Then we make the choice
\begin{equation}
f(x)=\frac{k a^\frac{1}{k}}{2\Gamma\left(\frac{1}{k}\right)}e^{-x^k},\;k=2,4,\ldots
\label{eq:def_f}
\end{equation}
duly normalized to unity, as prior density, The optimization of the upper bound (\ref{eq:S_le_rholnf}) for the choice (\ref{eq:def_f}) with respect to the parameter $a$ yields the following optimal bound
\[
S[\rho_n]\le \ln\left[A_k\langle x^k\rangle_n^\frac{1}{k}\right],\;k=2,4,\ldots
\]
with the constant
\[
A_k=\frac{2(ek)^\frac{1}{k}}{k}\Gamma\left(\frac{1}{k}\right).
\]

Finally, taking into account the expectation values $\langle x^k\rangle_n$ of the Rakhmanov density $\rho_n$ of the Hermite polynomial $\tilde{H}_n(x)$ given by Eq. (\ref{eq:xk}), we have that
\begin{eqnarray*}
S[\rho_n]\le \ln\left[\frac{(ek)^\frac{1}{k}}{k}\Gamma\left(\frac{1}{k}\right)
\left[\frac{k!}{\Gamma\left(\frac{k}{2}+1\right)}\,_2F_1\left(\left.
\begin{array}{c}
-n,-\frac{k}{2}\\
1
\end{array}
\right| 2
\right)\right]^\frac{1}{k}\right],
\end{eqnarray*}
and
\begin{eqnarray}
N[\rho_n]\le \left[\frac{(ek)^\frac{1}{k}}{k}\Gamma\left(\frac{1}{k}\right)
\left[\frac{k!}{\Gamma\left(\frac{k}{2}+1\right)}\,_2F_1\left(\left.
\begin{array}{c}
-n,-\frac{k}{2}\\
1
\end{array}
\right| 2
\right)\right]^\frac{1}{k}\right]\equiv c_{k,n},
\label{eq:def_ckn}
\end{eqnarray}
as optimal bounds for the Shannon entropy and Shannon length of $\tilde{H}_n(x)$, respectively. In Table \ref{tab:ckn} we give the value $k_{\rm opt}$ of $k$ which provides the best upper bound $c_{k,n}$ for the first few values of $n$.

\begin{table}
\begin{center}
\begin{tabular}{|c|ccccccccccccc|}
\hline
$n$ & 0&1&2&3&4&5&6&7&8&9&10&11&12\\
\hline
$k_{\rm opt}$ & 2&6& 8& 10& 12& 14& 16& 16& 18& 20& 22& 22& 24\\
\hline
$c_{k,n}$ & 2.92&4.54&5.57&6.40&7.11&7.75&8.33&8.86&9.36&9.83&10.30&10.70&11.10\\
\hline
\end{tabular}
\caption{Values of $k$ which provides the best upper bound $c_{k,n}$ for the first few values of $n$.}
\label{tab:ckn}
\end{center}
\end{table}

\subsection{Fisher length}

Finally, let us point out that the Fisher information of the Hermite polynomials $\tilde{H}_n(x)$ is given \cite{sanchezruiz:jcam05} by
\begin{equation}
F[\rho_n]=\int_{-\infty}^{+\infty} \left\{\frac{d}{dx}\rho_n(x)\right\}^2\frac{dx}{\rho_n(x)}=4n+2.
\label{eq:fisher_H}
\end{equation}
Thus, the Fisher length of these polynomials is
\begin{equation}
(\delta x)_n=\frac{1}{\sqrt{F[\rho_n]}}=\frac{1}{\sqrt{4n+2}},
\label{eq:fisher_length_rhon}
\end{equation}
whose values lie down within the interval $[0,\frac{1}{\sqrt{2}}]$. The comparison of this expression with Eq. (\ref{eq:Delta_x}) allows us to point out that not only the general relation (\ref{eq:delta_inequalities}) between the Fisher length and the standard deviation is fulfilled, but also that $(\delta x)_n(\Delta x)_n=\frac12$.

\section{Effective computation of the spreading lengths}
\label{sec4}

In this section we examine and discuss some computational issues relative to the spreading lengths of Hermite polynomials $H_n(x)$ and their mutual relationships. In contrast with the polynomials orthogonal on a segment of the real axis for which an efficient algorithm based on the three-term recurrence relation has been recently discovered for the computation of the Shannon entropy by V. Buyarov et al. \cite{buyarov:sjsc04}, up until now there is no explicit formula or at least a stable numerical algorithm for the computation of the Renyi and Shannon lengths of unbounded orthogonal polynomials, such as Hermite polynomials, for any reasonable $n\in\mathbb{N}$. A naive numerical evaluation of these Hermite functionals by means of quadratures is not often convenient except for the lowest-order polynomials since the increasing number of integrable singularities spoils any attempt to achieve a reasonable accuracy for arbitrary $n$.

We propose an analytical, error-free and easily programmable computing approach for the entropic moments $W_q[\rho_n]$ and the Renyi spreading lengths $L_q^R[\rho_n]$ of the orthonormal polynomials $\tilde{H}_n(x)$, which are given by Eqs. (\ref{eq:wqrhon_bell}) and (\ref{eq:lqrhon}), respectively, in terms of the combinatorial multivariable Bell polynomials defined by Eq. (\ref{eq:bell_def}). This approach requires the knowledge of the expansion coefficients $c_l$ given by Eq. (\ref{eq:def_c_l}) and the determination of the partitions $\tilde{\pi}(m,l)$ given by Eq. (\ref{eq:partitions}).

\begin{figure}
\begin{center}
\includegraphics[width=10cm]{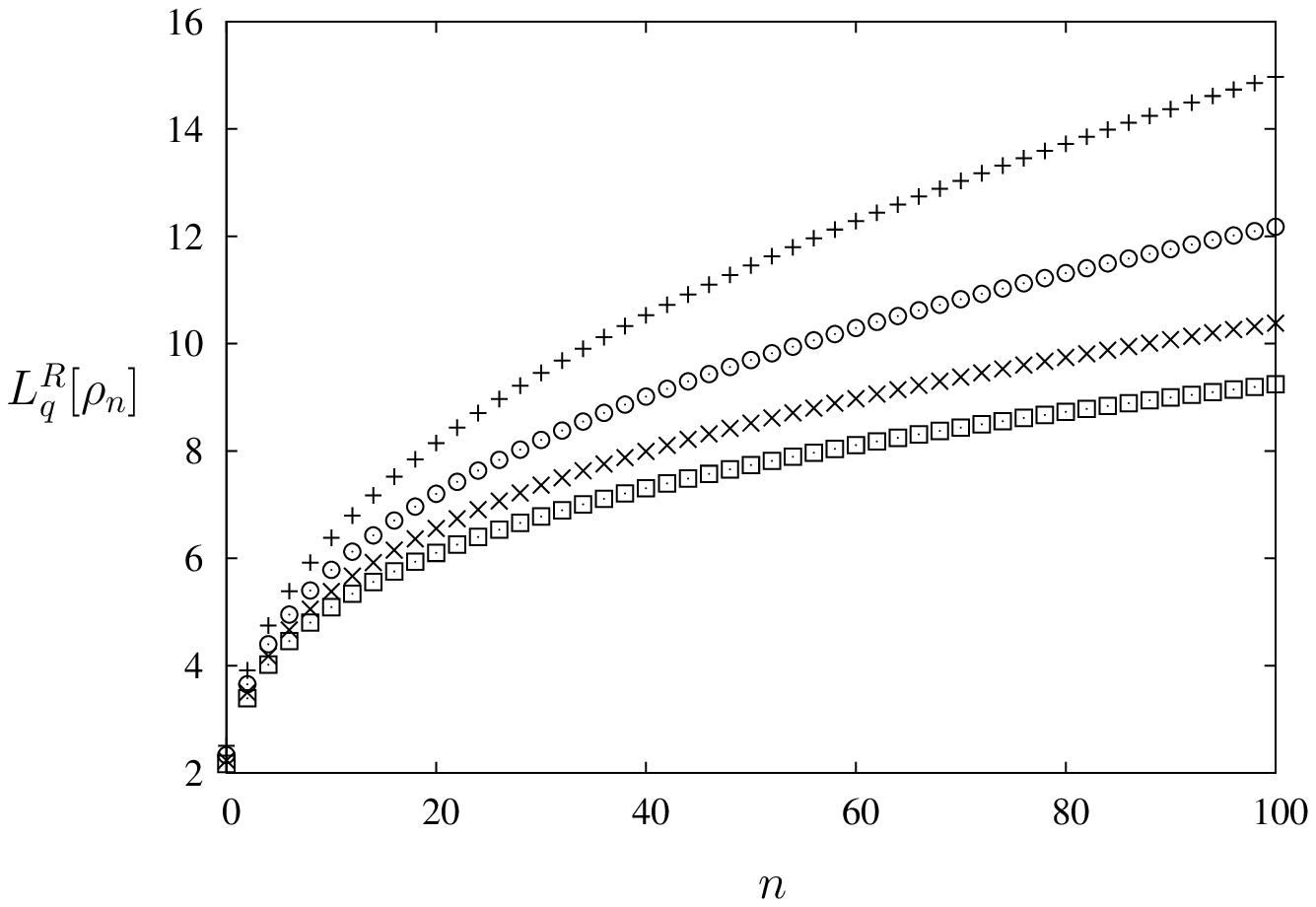}
\end{center}
\caption{Renyi lengths $L_q^R[\rho_n]$ with $q=2$ ($+$), $3$ ($\odot$), $4$ ($\times$), and $5$ ($\boxdot$), in terms of the degree $n$.}
\label{fig:1}
\end{figure}

\begin{figure}
\begin{center}
\includegraphics[width=10cm]{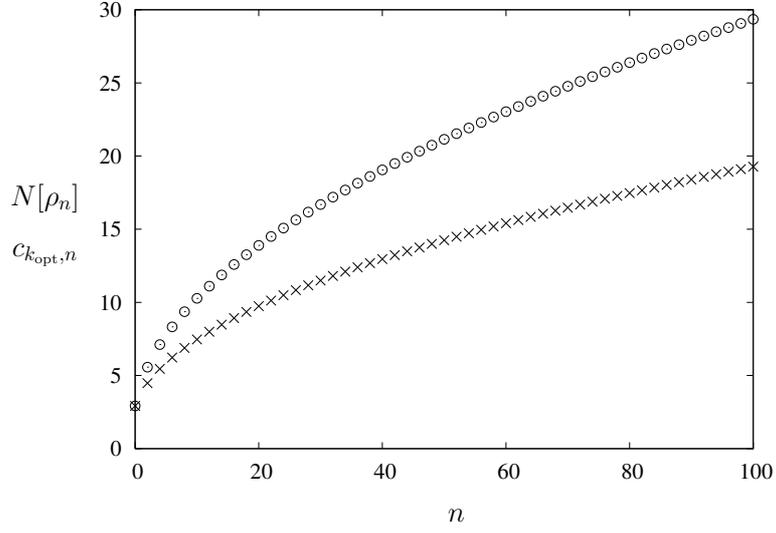}
\end{center}
\caption{Shannon length $N[\rho_n]$ ($\times$) and its optimal upper bound $c_{k_{\rm opt},n}$ ($\odot$) in terms of the degree $n$.}
\label{fig:2}
\end{figure}

In Figures \ref{fig:1} and \ref{fig:2} we have shown the Renyi lengths $L_q^R[\rho_n]$ with $q=2,3,4$ and $5$, and the Shannon length $N[\rho_n]$ of the orthonormal Hermite polynomials in terms of the degree $n$ within the range $n\in[0,100]$, respectively. We observe that both Shannon and Renyi lengths with fixed $q$ have an increasing parabolic dependence on the degree $n$ of the polynomials. Moreover, this behaviour is such that for fixed $n$ the Renyi length decreases when $q$ is increasing. As well, in Figure \ref{fig:2} we plot the optimal upper bound $c_{k_{\rm opt},n}$ for the Shannon length $N[\rho_n]$ according to Eq. (\ref{eq:def_ckn}).

\begin{figure}
\begin{center}
\includegraphics[width=10cm]{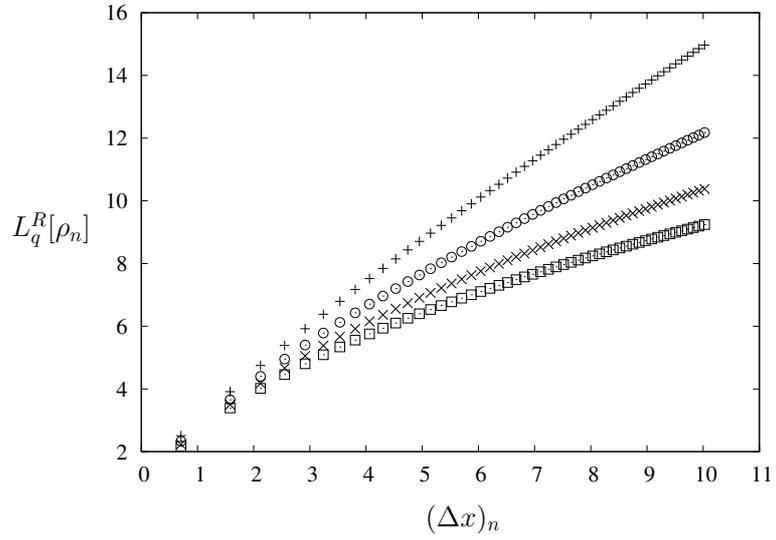}
\end{center}
\caption{Renyi lengths $L_q^R[\rho_n]$ with $q=2$ ($+$), $3$ ($\odot$), $4$ ($\times$), and $5$ ($\boxdot$), in terms of the standard deviation $(\Delta x)_n$ for $n\in[0,100]$.}
\label{fig:3}
\end{figure}

\begin{figure}
\begin{center}
\includegraphics[width=10cm]{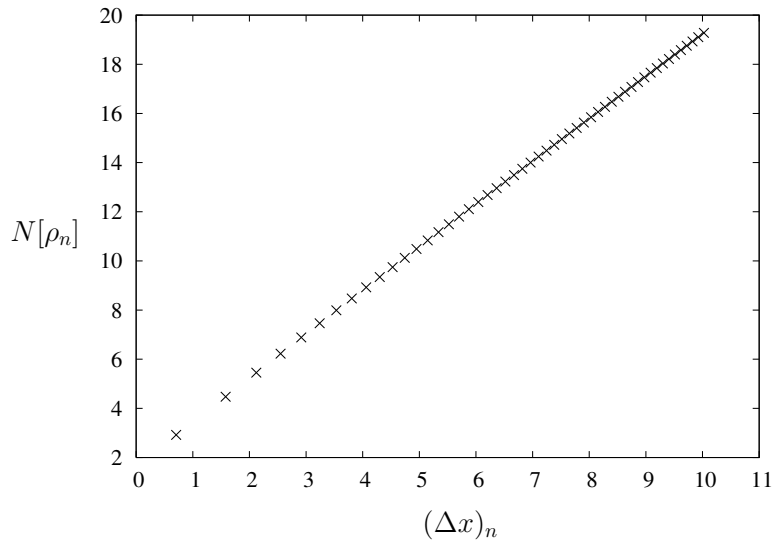}
\end{center}
\caption{Shannon length $N[\rho_n]$ in terms of the standard deviation $(\Delta x)_n$ for $n\in[0,100]$.}
\label{fig:4}
\end{figure}

Finally, for completeness, we have numerically studied the relation of the Shannon length $N[\rho_n]$ and the Onicescu-Heller length $L[\rho_n]$ of these polynomials with the standard deviation $(\Delta x)_n$ within the range $n\in[0,100]$. The corresponding results are shown in Figures \ref{fig:3} and \ref{fig:4}, respectively. We observe that when $n$ varies from $0$ to $100$ the two spreading lengths have a quasilinear behaviour in terms of the standard deviation. Moreover, we found the fit
\begin{equation}
N[\rho_n]=1.723 (\Delta x)_n+2.00
\label{eq:fitN}
\end{equation}
for the Shannon length, with a correlation coefficient $R=0.999998$, and
\[
L[\rho_n]=1.204(\Delta x)_n+2.92,
\]
for the Onicescu-Heller length or second-order Renyi length with a correlation coefficient $R=0.99994$. In fact, the global dependence of $N[\rho_n]$ on $(\Delta x)_n$ is slightly concave, being only asymptotically linear according to the rigorous expression (\ref{eq:nrho_delta}), i.e. $N[\rho_n]\simeq \frac{\pi\sqrt{2}}{e}(\Delta x)_n\simeq 1.63445 (\Delta x)_n$ for $n\gg 1$, which is in accordance with (\ref{eq:fitN}). It is worthwhile remarking (i) the $n^\alpha$ behaviour with $\alpha>0$ of the global spreading lengths (standard deviation, Renyi and Shannon lengths) and (ii) the $n^{-\frac{1}{2}}$-law which is followed by the (local) Fisher length. This difference may be associated to the gradient-functional form (\ref{eq:def_deltax}) of the Fisher length, indicating a locality property in the sense that it is very sensitive to the oscillatory character of the polynomials.

\section{Application to the harmonic oscillator}
\label{sec5}

The states of the one-dimensional harmonic oscillator, with a potential $V(x)=\frac12 \lambda^2 x^2$, are determined by the following probability density 
\[
\rho_n^{\rm HO}(x)=\frac{\sqrt{\lambda}}{2^nn!\sqrt{\pi}}e^{-\lambda x^2} \left(H_n(\sqrt{\lambda}x)\right)^2
\]
where $H_n(x)$ are the orthogonal Hermite polynomials \cite{nikiforov_88}. Since $\tilde{H}_n(x)=(2^nn!\sqrt{\pi})^{-\frac12} H_n(x)$, this density can be expressed in terms of the Rakhmanov density of the orthonormal Hermite polynomials $\tilde{H}_n(x)$ defined by Eq. (\ref{eq:rho_n_hermite}) as follows:
\[
\rho_n^{\rm HO}(x)=\sqrt{\lambda} \rho_n(\sqrt{\lambda} x).
\]
Taking this into account, we obtain for the harmonic oscillator the following expressions
\[
\langle x^k\rangle_n^{\rm HO}=\lambda^{-\frac{k}{2}}\langle x^k\rangle_n;\,k=0,1,2,\ldots
\]
\[
W_q[\rho_n^{\rm HO}]=\lambda^{\frac{q-1}{2}}W_q[\rho_n];\,q=0,1,2,\ldots
\]
for the ordinary and entropic moments,
and
\[
S[\rho_n^{\rm HO}]=-\ln\sqrt{\lambda}+S[\rho_n],
\]
\[
F[\rho_n^{\rm HO}]=\lambda F[\rho_n],
\]
for the Shannon and Fisher information measures
in terms of the corresponding quantities of the Hermite polynomials, whose values are given by Eqs. (\ref{eq:xk}), (\ref{eq:wqrhon_bell}), (\ref{eq:shannon_H}) and (\ref{eq:fisher_H}), respectively. Then, it is straightforward to find that all the uncertainty measures of the harmonic oscillator (namely, standard deviation and Renyi, Shannon and Fisher lengths) are equal, save for a multiplication factor $\lambda^{-\frac12}$, to the corresponding spreading lengths of the Hermite polynomials which control their wavefunctions. This is to say that
\[
(\Delta x)_n^{\rm HO}=\lambda^{-\frac12}(\Delta x)_n, \quad L_q^R[\rho_n^{\rm HO}]=\lambda^{-\frac12} L_q^R[\rho_n]
\]
\[
N[\rho_n^{\rm HO}]=\lambda^{-\frac12}N[\rho_n],\quad (\delta x)_n^{\rm HO}=\lambda^{-\frac12}(\delta x)_n,
\]
whose values can be obtained by keeping in mind Eqs. (\ref{eq:Delta_x}), (\ref{eq:lqrhon})-(\ref{eq:shannon_H}), and (\ref{eq:fisher_length_rhon})  respectively.

\section{Open problems}

Here we want to pose the following unsolved problems: to find the asymptotics of the frequency or entropic moments $W_q[\rho_n]$ defined by
\begin{equation}
W_q[\rho_n]=\int_{-\infty}^\infty \left[\omega_H(x) \tilde{H}_n^2(x)\right]^qdx = \int_{-\infty}^\infty e^{-qx^2}\left[\tilde{H}_n(x)\right]^{2q}dx,\,q\ge 1
\label{eq:wqrhon_2}
\end{equation}
in the two following cases:
\begin{eqnarray}
\begin{array}{l}
\bullet\; n\to\infty, \text{ and } q\in\mathbb{R} \text{ fixed}.\\
\bullet\; q\to\infty, \text{ and } n\in\mathbb{N} \text{ fixed}.
\end{array}
\label{eq:problems}
\end{eqnarray}

Up until now the only known result is due to Alexander Aptekarev et al \cite{aptekarev:rassm95,aptekarev:jcam09,aptekarev:dm96} who have shown that these quantities, which are weighted $L^{2q}$-norms of the orthonormal Hermite polynomials $\tilde{H}_n(x)$ as
\[
W_q[\rho_n]=\parallel\tilde{H}_n(x)\parallel_{2q}^{2q},
\]
behave asymptotically ($n\to\infty$) as
\[
W_q[\rho_n]=\left(\frac{2}{\pi}\right)^q\frac{\Gamma\left(q+\frac12\right)\Gamma\left(1-\frac{q}{2}\right)}{\Gamma(q+1)\Gamma\left(\frac32-\frac{q}{2}\right)}(2n+1)^\frac{1-q}{2}(1+o(1))
\]
when $q\in[0,\frac43]$. To extend and generalize this result in the sense mentioned above, it might be useful for the interested reader to keep in mind the two following related results:
\begin{itemize}
\item[(a)] Lars Larsson-Cohn \cite{larsson:am02} studied the $L^p$-norms of the monic Hermite polynomials $h_n(x)$, orthogonal with respect to the Gaussian weight $\omega_G(x)=(2\pi)^{-\frac12}\exp\left(-\frac{x^2}{2}\right)$, defined by
\[
\|h\|_p=\left(\frac{1}{2\pi}\right)^\frac{p}{2}\left\{\int_{-\infty}^\infty e^{-\frac{x^2}{2}}|h_n(x)|^pdx\right\}^\frac{1}{p}.
\]
Keeping in mind that $\|h\|_n=\sqrt{n!}$, he found the following asymptotical $(n\to\infty)$ result
\begin{eqnarray*}
\|h\|_p=\left\{
\begin{array}{ll}
c(p)n^{-\frac14}\sqrt{n!}(1+O(n^{-1})), & \text{if } 0<p<2\\
c(p)n^{-\frac14}\sqrt{n!}(p-1)^\frac{n}{2}(1+O(n^{-1})), & \text{if } 2<p<\infty
\end{array}
\right.
\end{eqnarray*}
with the values
\begin{eqnarray*}
c(p)=\left\{
\begin{array}{ll}
\left(\frac{2}{\pi}\right)^\frac14\left(\frac{2}{2-p}\right)^\frac{1}{2p}\left\{\frac{\Gamma\left(\frac{p+1}{2}\right)}{\sqrt{\pi}\Gamma\left(\frac{p+2}{2}\right)}\right\}^\frac{1}{p}, &\text{for } p<2\\
\left(\frac{2}{\pi}\right)^\frac14\left(\frac{p-1}{2p-4}\right)^\frac{p-1}{2p}, & \text{for } p>2
\end{array}
\right.
\end{eqnarray*}

However, the asymptotics of the entropic moments $W_q[h_n]$ of the Hermite polynomials $h_n(x)$ defined by
\begin{equation}
W_q[h_n]=\int_{-\infty}^\infty \left[\omega_G(x) h_n^2(x)\right]^q dx =\left(\frac{1}{2\pi}\right)^\frac{q}{2}\int_{-\infty}^\infty e^{-\frac{q}{2}x^2}\left[h_n(x)\right]^{2q} dx
\label{eq:wqhn_larsson}
\end{equation}
has not yet been found.

\item[(b)] Ruth Azor et al \cite{azor:sjma82} have used combinatorial techniques for the asymptotics ($n\to\infty$) of the following functionals of the Hermite polynomials $H_n(x)$ orthogonal with respect to $e^{-x^2}$:
\[
Z_k[H_n]:=\int_{-\infty}^\infty e^{-x^2}[H_n(x)]^k dx, k\in\mathbb{N}.
\]
They were only able to solve it in the particular non-trivial case $k=4$, where
\[
Z_4[H_n]=\frac{3}{4n}\sqrt{\frac{3}{\pi}}\frac{6^{2n}}{(n!)^2}\left\{1-\frac{1}{4n}+\frac{3}{16n^2}+O(n^{-3})\right\}.
\]
Besides, they have also considered the asymptotics ($k\to\infty$) of $Z_k[H_n]$ as well as that of the quantities
\[
D_k[H_n]:=\sqrt{\frac{2}{\pi}}\int_{-\infty}^\infty e^{-2x^2} \left[\tilde{H}_n(x)\right]^kdx,
\]
finding the following results
\[
Z_k[H_n]\sim \left[1+(-1)^{kn}\right]\left(2^{kn-1}\pi\right)^\frac12 \left(\frac{kn}{e}\right)^\frac{kn}{2} e^{-\frac{n-1}{2}},\;\text{for } k\gg 1
\]
\[
D_k[H_n]\sim\frac{1+(-1)^{kn}}{\sqrt{2}}\left(\frac{kn}{e}\right)^{\frac{kn}{2}}e^{-(n-1)},\;\text{for } k\gg 1
\]
\end{itemize}

The solution of the problems (\ref{eq:wqrhon_2})-(\ref{eq:problems}) and/or (\ref{eq:wqhn_larsson}), and the full determination of the asymptotics of $Z_k[H_n]$ and $D_k[H_n]$ for ($n\to\infty$, $k\in\mathbb{R}$),
either by means of an approximation-theoretic methodology \cite{aptekarev:rassm95,aptekarev:jcam09,aptekarev:dm96,larsson:am02} or by use of combinatorial techniques \cite{azor:sjma82,desainte:lnm85}, is of great relevance from both mathematical and applied points of view from obvious reasons, keeping in mind the physico-mathematical meaning of these quantities as previously mentioned. It is worthwhile remarking that, in particular, the quantity $W_q[\rho_n]$ is not only the $(2q)$th-power of the $L^{2q}$-norm of the Hermite polynomials but it also represents the entropic moment of order $q$ of the harmonic oscillator and their isospectral physical systems, which fully determines the Renyi and Tsallis entropies of these objects.

\section{Conclusions}

In this paper we have introduced new direct measures of orthogonal-polynomials spreading other than the root-mean-square or standard deviation $\Delta x$ which have an information-theoretic origin; namely, the Renyi, Shannon and Fisher length. They share with $\Delta x$ various interesting properties: same units as the variable, invariance under translations and reflections, linear scaling, vanishing in the limit that the variable tends towards a given definite value, and global character. In contrast with $\Delta x$, they do not depend on any particular point of the orthogonality interval what allow them to be considered as proper spreading lengths.

The Renyi and Shannon lengths are powerlike and logarithmic functionals of the polynomial $p_n(x)$ while Fisher length is a gradient functional of it, what allows one to state that the former lengths are global measures and the latter one has a locality property. For the Renyi length of the orthonormal Hermite polynomials $\tilde{H}_n(x)$ we have developed a computational methodology which is based on the combinatorial multivariable Bell polynomials whose arguments are controlled by the expansion coefficients of $\tilde{H}_n(x)$. For the Shannon length of these polynomials, since it cannot be calculated explicitly, we give (i) its asymptotics and the relation with the standard deviation, and (ii) sharp upper bounds by means of an information-theoretic-based optimization procedure. Moreover, it is computationally found that the second-order Renyi length (also called Onicescu-Heller length) and the Shannon length linearly depend on the standard deviation of the polynomials. On the other hand, the Fisher length is analytically shown to have a reciprocal behaviour with $\Delta x$ mainly because of its local character.

Finally, the previous results have been applied to the quantum-mechanical harmonic oscillator and some open problems related to the asymptotics of the frequency or entropic moments of Hermite polynomials are posed. Their solution would allow not only to complete this work but also to extend considerably the related findings of various authors in connection to the $L^p$-norms of Hermite polynomials \cite{aptekarev:rassm95,aptekarev:jcam09,aptekarev:dm96,azor:sjma82,larsson:am02}.

\appendix
\section{$p$th-power of a polynomial of degree $n$ and Bell polynomials}

Consider the polynomial $y_n(x)$ with degree $n$,
\[
y_n(x)=\sum_{k=0}^n c_k x^k.
\]
Its $p$th-power is
\[
[y_n(x)]^p=\left(\sum_{k=0}^n c_k x^k\right)^p =\sum_{\pi(p)}\frac{p!}{j_0! j_1!\cdots j_n!}(c_0x^0)^{j_0} (c_1x^1)^{j_1}\cdots(c_nx^n)^{j_n},
\]
where the sum is defined over all the partitions $\pi(p)$ such that $j_0+j_1+\cdots +j_n=p$.

The previous expression of $[y_n(x)]^p$ can be written as
\[
[y_n(x)]^p=\sum_{\pi(p)}\frac{p!}{j_0! j_1!\cdots j_n!} c_0^{j_0} c_1^{j_1}\cdots c_n^{j_n} x^{j_1+2j_2+\cdots+nj_n}
=\sum_{k=0}^{np} A_{k,p}(c_0,\ldots,c_n) x^k,
\]
where
\[
A_{k,p}(c_0,\ldots,c_n)=\sum_{\tilde{\pi}(k,p)}\frac{p!}{j_0! j_1!\cdots j_n!} c_0^{j_0} c_1^{j_1}\cdots c_n^{j_n}
\]
where the sum is defined over all the partitions $\tilde{\pi}(k,p)$ such that
\[
j_0+j_1+\cdots+j_n=p,\quad \text{and}\quad j_1+2j_2+\cdots+nj_n=k.
\]

The Bell polynomials are defined as
\[
B_{m,l}(c_1,c_2,\ldots,c_{m-l+1})=\sum_{\hat{\pi}(m,l)} \frac{m!}{j_1! j_2!\cdots j_{m-l+1}!}\left(\frac{c_1}{1!}\right)^{j_1} \left(\frac{c_2}{2!}\right)^{j_2}\cdots \left(\frac{c_{m-l+1}}{(m-l+1)!}\right)^{j_{m-l+1}}
\]
where the sum is defined over all the partitions $\hat{\pi}(m,l)$ such that
\[
j_1+j_2+\cdots+j_{m-l+1}=l,\quad\text{and}\quad j_1+2j_2+\cdots+(m-l+1)j_{m-l+1}=m.
\]

The relation between the coefficients $A_{k,p}(c_0,\ldots, c_{k-p+1})$ and the Bell polynomials can be established as
\[
A_{k,p}(c_0,\ldots,c_n)=\frac{p!}{k!}\sum_{j_0=0}^p \frac{c_0^{j_0}}{j_0!} B_{k,p-j_0}(c_1,2!c_2,\ldots,(k-p+j_0+1)!c_{k-p+j_0+1}),
\]
where $c_i=0$ for $i>n$.

From Equation (3l) from \cite{comtet_74} we obtain that
\[
\frac{1}{k!}\sum_{j_0=0}^p \frac{c_0^{j_0}}{j_0!} B_{k,p-j_0}(c_1,2!c_2,\ldots,(k-p+j_0+1)!c_{k-p+j_0+1})
=\frac{1}{(k+p)!} B_{k+p,p}(c_0,2!c_1,\ldots,(k+1)!c_{k}).
\]
Then,
\[
A_{k,p}(c_0,\ldots,c_n)=\frac{p!}{(k+p)!}B_{k+p,p}(c_0,2!c_1,\ldots,(k+1)!c_{k}),
\]
where, again, $c_i=0$ for $i>n$.

\section*{Acknowledgements}

We are very grateful for partial support to Junta de Andaluc\'{\i}a
(under grants FQM-1735 and FQM-2445) and Ministerio de
Educaci\'on y Ciencia under project FIS2008-2380. The authors belong to the Andalusian research group FQM-0207.

\end{document}